\documentclass[final]{ias2}

\usepackage{graphicx} 
\usepackage{multirow}
\usepackage{array} 

\usepackage{hyperref} 

\begin{document}

\markboth{Multiple time-scales in hierarchical modular networks}
{S. Sinha and S. Poria}

\title{Multiple dynamical time-scales in networks with 
hierarchically nested modular organization}

\author[sin,ysin]{Sitabhra Sinha} 
\author[ain]{Swarup Poria} 
\address[sin]{The Institute of Mathematical Sciences, CIT Campus,
Taramani, Chennai 600113, India}
\address[ain]{Department of Applied Mathematics, University of
Calcutta, 92 Acharya Prafulla Chandra Road, Kolkata 700009, India}
\address[ysin]{Corresponding author: sitabhra@imsc.res.in}

\begin{abstract}
Many natural and engineered complex networks have intricate
mesoscopic organization, e.g., the clustering of the constituent nodes
into several communities or modules. Often, such modularity is
manifested at several different hierarchical levels, where the
clusters defined at one level appear as elementary entities at the
next higher level. Using a simple model of a hierarchical modular
network, we show that such a topological structure gives rise to
characteristic time-scale separation between dynamics occurring at
different levels of the hierarchy. This generalizes our earlier result
for simple modular networks, where fast
intra-modular and slow inter-modular processes were clearly
distinguished. Investigating the process of synchronization of
oscillators in a hierarchical modular network, we show the
existence of as many distinct time-scales as there are hierarchical levels
in the system. 
This suggests a possible functional role of such
mesoscopic organization principle in natural systems, viz., in the dynamical
separation of events occurring at different spatial scales.
\end{abstract}


 
\maketitle


\section{Introduction}
Complex networks are ubiquitous~\cite{Newman10}. 
They are seen in phenomena studied by
physicists from
resistor networks to polymer contact structure to spin interactions in
disordered systems~\cite{Dasgupta09}, in engineered systems (e.g., the internet and the power
grid) and in the biological world from the intra-cellular signaling system to
neuronal networks to ecological food webs~\cite{Sinha09a}. Analysis of such networks
can lead not only to fundamental understanding of how complex systems
function, but may also have immediate practical relevance. For
example, understanding the network of intra-cellular communication
leading to identification of the key players can result in developing drugs targeted specifically towards these
molecules~\cite{Sinha09b}. On a much larger scale, by revealing how the contact
structure among susceptible and infected individuals govern the
propagation of epidemics, the manipulation of networks may play a
significant role in public health. 

Beginning in 1998-99, over the last decade interest in complex
networks
has grown by leaps and bounds among physicists. However, the focus of
research activity related to such systems has been on
either the microscopic properties of
individual nodes such as the distribution of degree (i.e., the number
of links for a node)~\cite{Barabasi99} or the macroscopic properties characterizing the
entire network in terms of a global value, such as average path length
or clustering coefficient~\cite{Watts98}. 
Recent research, on the other hand, has
revealed that networks which are indistinguishable at either of these
two scales may nevertheless have radically different
behavior~\cite{Pan09}. The
origin of this difference lies in their mesoscopic organization which
can be structurally manifested as patterns in the arrangement of links
between subparts of the network~\cite{Pan08}. 

One of the prominent examples of such organizing principles
operating in networks is the existence of communities, also referred
to as {\em modularity}~\cite{Newman06}. Modules can be defined as
subnetworks whose
components are much more densely and/or strongly connected with each
other as compared to connections with components that belong to
different subnetworks. 
In earlier work we have shown that existence of modules can give rise to two
distinct time-scales for the dynamics on such networks: a strongly
modular character of the system, as indicated by a low value for the
ratio of within-module to between-module connections, results in a
sharp distinction between fast intra-modular and slower
inter-modular processes~\cite{Pan09}.
Another mesoscopic organizing principle observed in many networks
is {\em hierarchy}, i.e., an arrangement in which entities are
ordered in several levels or layers~\cite{Simon62}. 
We are most familiar with hierarchy as seen in social
organization; however, they are also present in other systems, as for
example in the neuronal communication network which has a clearly
defined direction of information flow, beginning with the sensory
organs and culminating with motor response. While both modular and
hierarchical features are increasingly being reported in a wide range
of systems,
it is still not clear what role such patterns play in the
dynamics and function of the corresponding systems. Thus, the second
wave of research in complex networks that is just beginning focuses on
these mesoscopic aspects of networks. 

There can be even more complicated intermediate-scale
features in networks, in particular, hierarchically nested
modularity,
where modules and meta-modules (composed of multiple strongly
connected modules) may occur at different hierarchical levels. 
Such hierarchical modularity has been observed in many naturally
occurring systems, including biological metabolic networks~\cite{Ravasz02}, 
ecological food webs~\cite{Bastolla09},
social groups~\cite{Zhou05,Hamilton07}, financial markets~\cite{Leibon08} and
brain functional networks~\cite{Zhou07,Bassett10}.
Fig.~\ref{fig:fig1}~(a-b) shows empirical evidence for the
occurrence of hierarchical modular
organization in the network of connections between cortical regions in the
cat~\cite{Scannell95} and
macaque~\cite{Honey07} brains obtained from anatomical studies. 
Theoretical understanding of such systems is still at an early stage. A
simple-minded deterministic construction approach towards such
networks claimed that such
structures have a specific signature in the power law dependence of
the clustering coefficient of nodes having degree $k$ on
$k$~\cite{Ravasz03}. We 
demonstrated, using simple counter-examples, that this is true only for
a very specific set of artificially generated networks and does not
hold in general~\cite{Pan08}. This illustrates the importance of
acquiring a
clearer understanding of the properties that the hierarchical nesting
of modular structures impart to a system. We focus specifically on
the role that such mesoscopic organization plays in synchronization
dynamics on the network.

In this paper we show that in a hierarchical modular network of
oscillators arranged into
$l = h_{lev}$ levels, there exist $h_{lev}$ distinct time-scales of the
synchronization dynamics. The oscillators belonging to the elementary
modules ($l = 0$) synchronize earliest, followed by those belonging to the same
meta-module ($l = 1$) and so on, with global synchronization being
achieved last~\footnote{Note that if the synchronization time diverges
rapidly, we may not observe synchronization beyond a certain level
within a reasonable period of observation and thus, global
synchronization may never be achieved.}.
In the next section we discuss the simple network model with
hierarchically nested modules which we have used.
Our results on the synchronization dynamics on such a
network are described in Section 3. We have carried out spectral
analysis of the Laplacian
matrix for the network to give a theoretical background for the explicit
numerical simulations. We conclude with a short discussion on the
possible functional relevance for the separation of time-scales
through hierarchical modular organization in the biological world.

\section{The Model}
\begin{figure*}
\begin{center}
\includegraphics[width=0.99\textwidth]{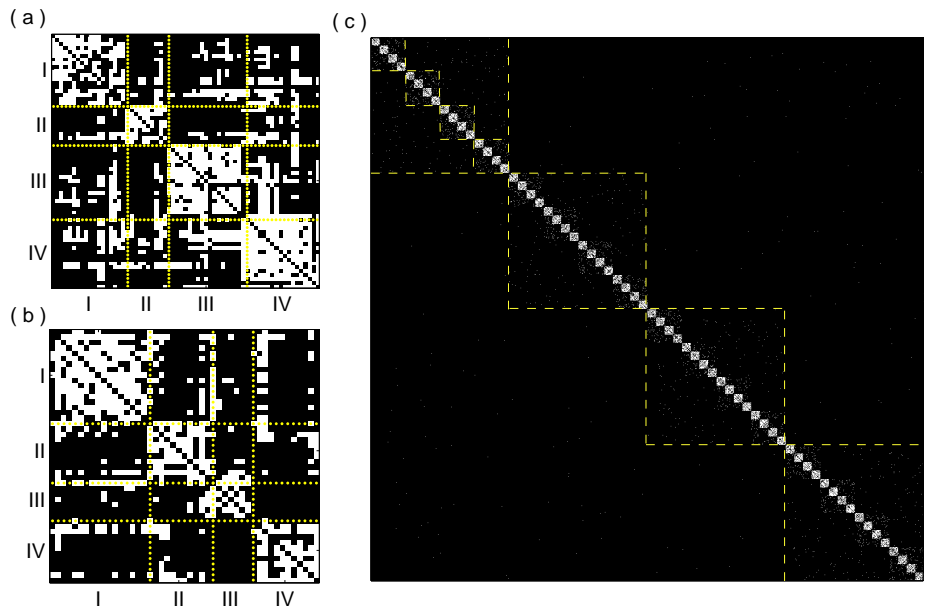}
\caption{The adjacency matrix corresponding to the network of
cortico-cortical connections in the (a) cat and (b) macaque brains.
The largest-scale modules are indicated by dotted lines and labeled
with Roman numerals.
(c) Adjacency matrix for the hierarchical modular network of
$N=1024$ nodes with average degree $\langle k \rangle$. There are
$h_{lev}=3$ hierarchical levels, with a branching ratio of $q=4$ at
each level. The ratio of connection density between two successive
hierarchical levels, $r=0.05$. The modules at each hierarchical level
are indicated by broken lines.
}
\label{fig:fig1}
\end{center}
\end{figure*}

We have used a general model for networks having modular organization
at several hierarchically arranged levels that has been introduced in
Ref.~\cite{Pan08} [Fig.~\ref{fig:fig1}~(c)]. At the lowest level ($l =
0$), the network consists of $M$ modules, each containing $n$ nodes.
The connectivity, or, the probability of a link, between a pair of
nodes is much higher for nodes belonging to the same module compared
to those belonging to different modules. Let us assume the
connectivity within each of the modules to be $0 < \rho_1 \leq 1$. 
We now introduce the next hierarchical level ($l = 1$) by grouping
together $m$ modules into ``meta-modules". The connection density
between nodes belonging to different modules in the same meta-module
is $\rho_2 (\leq \rho_1)$. Thus, while the oscillators in the different
modules within each of the $M/m$
meta-modules have a lower probability of connection ($\rho_2$) between them as
compared to the probability of connections between oscillators in the
same module ($\rho_1$), it is higher than the connectivity for oscillators
belonging to different meta-modules. To increase the number of
hierarchical levels to three, we group $q$ such meta-modules into
``meta-meta-modules" ($l = 2$). The connectivity between oscillators
belonging to different meta-modules but in the same meta-meta-module
is $\rho_3 (\leq \rho_2)$. This procedure of grouping together
$q$ clusters of oscillators at higher and higher levels of organization
can be continued upto a maximum number of layers, $h_{lev}$,
implicitly given by the relation
$q^{h_{lev}-1} m = M$. The parameter $q$ is the branching at each
level (except the lowest one), corresponding to the number of clusters
at one level that are grouped together to form the cluster at the next
higher level.

To reduce the number of model parameters, we further assume that the
connectivities within and between clusters at different levels,
$\rho_1, \rho_2, \ldots, \rho_{h_{lev}}$ are related as:
\begin{equation}
\frac{\rho_2}{\rho_1}=\frac{\rho_3}{\rho_2}=\cdots=\frac{\rho_{h_{lev}}}{\rho_{h_{lev}-1}}=r,
\end{equation}
where $0\leq r \leq 1$ is the ratio of inter-cluster connectivities at
two successive hierarchical levels. Varying $r$ from 0 to 1, one
obtains a whole range of hierarchical modular networks, with the
special case of $M$ isolated modules at one extreme ($r = 0$) and a
homogeneous Erdos-Renyi random network at the other ($r = 1$).
Figure~\ref{fig:fig1}~(c) shows an example of a hierarchical network with 3
levels and $r=0.05$ constructed following the above procedure.

The model allows the modular character of the network to be changed
(by varying $r$) independent of the hierarchical complexity which
is governed by the number of levels, $h_{lev}$. The stochastic
construction process of the network implies that reasonable
statistical averages can be done over an ensemble of
random networks with the same hierarchical modularity.

\section{Results}

Most networks occurring in nature have dynamics associated with their
nodes~\cite{Strogatz01}. In particular, many systems comprise
oscillating elements,
such as, pancreatic beta cells, neurons or cardiac pacemaker
cells~\cite{Goldbeter97,Winfree00}. 
Synchronization of the periodic activity exhibited by
the nodes in such networks may play a vital functional role. To
observe how the hierarchical modular organization of a network can
affect its synchronization behavior, we consider a population of $N$
coupled oscillators. The
time-evolution of the phase of
the $i$-th oscillator having a frequency $\omega_i$ and which is
connected to  $k_i$ other oscillators,
can be described by~\cite{Acebron05}:
\begin{equation}
\frac{d\theta_i}{dt} = \omega_i + \frac{1}{k_i}
\displaystyle\sum\limits_{j=1}^N ~K_{ij}~{\rm sin}(\theta_j -
\theta_i).
\label{eq:eq1}
\end{equation}
The coupling between the different oscillators is described by the
matrix {\bf K} = $\{ K_{ij} \} = \kappa${\bf A}, where $\kappa (=1)$ is the
strength of interaction between the oscillators 
and {\bf A} is the
network adjacency matrix, i.e., $A_{ij} = 1$ if nodes $i,j$ are
connected, and 0 otherwise. In our analysis, we will mostly consider
the simplified case of oscillators with identical frequency, i.e.,
$\omega_i = \omega$. However, we have explicitly verified that
randomly varying the frequencies of different oscillators
over a small range do not qualitatively affect our
results.

The attractor for the network dynamics described by
Eq.~\ref{eq:eq1} is the fully synchronized state $\theta_i = \theta$,
$\forall i$. However, for a network with mesoscopic organization, 
the convergence to global synchronization occurs in a series
of steps which is intrinsically related to the underlying connection
topology (and has been seen earlier in simple modular networks~\cite{Arenas06}).
We observe that, starting from a random distribution of initial phases
for the different oscillators, the nodes within each module
synchronize first, followed by larger and larger structures at the
higher hierarchical levels in sequence till the entire system
oscillates in the same phase. The distinct time-scales of
synchronization for the different levels reflect the structural
organization of the network.

\begin{figure*}
\begin{center}
\includegraphics[width=0.99\textwidth]{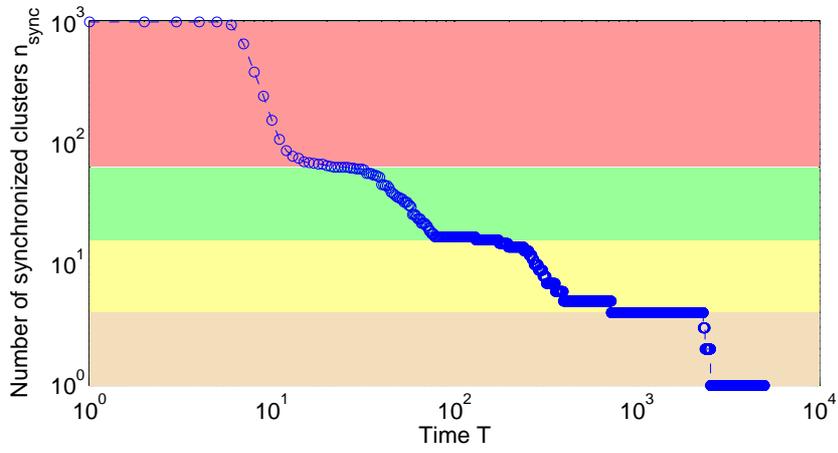}
\caption{Time-evolution of the number of synchronized clusters of
oscillators for hierarchical modular network with $N=1024$ nodes of
average degree $\langle k \rangle = 14$. The synchronization occurs
over three distinct time-scales, reflecting the number of hierarchical
levels $h_{lev}=3$ of the network. First, local synchronization
occurs by $T\simeq 20$ among the oscillators within each of the 64 
modules (having 16 oscillators each) at the
lowest level, $l=0$. By $T \simeq 80$, groups of four 
modules (corresponding to a total of 64
oscillators each), which form the meta-module in the next 
hierarchical level ($l=1$) get synchronized. By $T \simeq 720$,
synchronization is observed at the level $l = 2$, where
each of the synchronized groups comprise 256 oscillators. Finally
global synchronization among all 1024 oscillators occurs at $T \sim
2500$.
The branching at each level is
$q=4$ and the ratio of
intermodular connections between two successive hierarchical levels,
$r = 0.03$.}
\label{fig:fig2}
\end{center}
\end{figure*}

To analyze the time evolution of the synchronization process, we
measure a local order parameter using the pair-correlation function
between the oscillator phases:
\begin{equation}
\rho_{ij} (t) = \langle cos [\theta_i (t) - \theta_j (t)] \rangle,
\end{equation}
where $\langle \ldots \rangle$ is an average over random initial
phases. Introducing a threshold $T$, the correlation matrix is
converted into a dynamic connectivity matrix ${\cal D}_t (T)$, where
${\cal D}_{ij} = 1$ if $\rho_{ij} > T$ and $=0$ otherwise. The number of
distinct synchronized communities at any given time is given by the 
number of disconnected clusters in $\cal D$. Initially, when the
phases of the oscillators are chosen from a random distribution, none
of the elements are synchronized and the number of clusters is
trivially equal to the number of oscillators, $N$. However, with time,
more and more oscillators get successively synchronized and when
global synchronization is achieved, there is only a single cluster of
synchronized elements. In contrast to
homogeneous random networks, where there is a smooth, continuous 
transition to global synchronization with time, for a hierarchical
modular network we observe a series of step-like transitions as larger
and larger clusters get synchronized at different time-scales
(Fig.~\ref{fig:fig2}).

For a network having $h_{lev}$ hierarchical levels, the evolution to
global synchronization exhibits $h_{lev}$ intermediate time-scales as
reflected by the occurrence of $h_{lev}$ plateaus between $N$ and 1 for 
the number of synchronized clusters (Fig.~\ref{fig:fig2}). First, at the
relatively short time-scale of $\tau_m$, disconnected clusters are
observed to form in $\cal D$ which correspond to the modules at the
lowest hierarchical level of the network. Thus, local synchronization
within each module at level $l = 0$ is achieved relatively rapidly.
Next, at a slightly longer time-scale of $\tau_{mm}$, several of the clusters
mentioned earlier merge into bigger clusters corresponding to the
meta-modules at the next hierarchical level $\l = 1$ of the network.
Therefore, the intermediate-scale synchronization over meta-modules is
a slower process than the local synchronization seen within each
module, the initially synchronized clusters remaining fairly stable in
the intervening time-period between $\tau_m$ and $\tau_{mm}$.
Successive hierarchical levels take longer and longer
times to synchronize, and we can associate distinct synchronization
time-scales $\tau$ with each such level. Finally, global
synchronization is observed at the longest time-scale of $\tau_g$.

\begin{figure*}
\begin{center}
\includegraphics[width=0.8\textwidth]{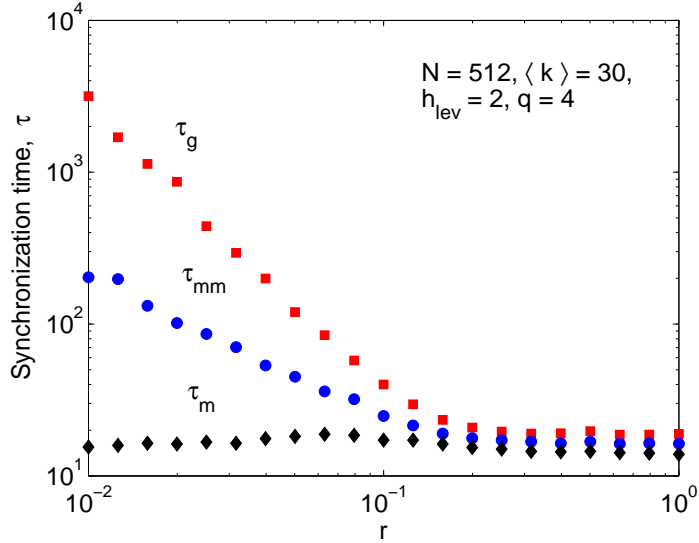}
\caption{The variation of the synchronization time-scales at different
hierarchical levels as a function of $r$ for a network having $N=512$
oscillators arranged into modules in $h_{lev}=2$ levels.
The time $\tau_m$ is that required for each module to synchronize the
32 oscillators that each of them have, while $\tau_{mm}$ is the time
required for synchronization within a meta-module comprising four of
the aforementioned modules and $\tau_g$ is the global synchronization time
(i.e., the entire system comprising four meta-modules). 
The three time-scales 
diverge sharply as the modularity at each level becomes more 
prominent with decreasing $r$. Conversely, as the network becomes more
homogeneous when $r \to 1$ the three time-scales converge.
The average
number of connections among the oscillators $\langle k \rangle = 30$
and the branching
at each level is $q=4$.}
\label{fig:fig3}
\end{center}
\end{figure*}

As the existence of the distinct time-scales to synchronization is a
result of the mesoscopic organization of the network, we expect that
as the system becomes more homogeneous on increasing $r$, the
different time-scales should converge towards the same value. This is
indeed observed in Fig.~\ref{fig:fig3}, where the three time-scales
for $h_{lev} = 2$, viz.,
$\tau_m$, $\tau_{mm}$ and $\tau_g$
approach each other as $r
\rightarrow 1$.

The existence of the distinct time-scales in a hierarchical modular
network can be understood by analyzing the linearized dynamics around
the synchronized state of the system:
\begin{equation}
\frac{d \theta_i}{dt} = \frac{1}{k_i} \sum_j L_{ij} \theta_j,
\label{eq:lap}
\end{equation}
where {\bf L} is the Laplacian of the network, with $L_{ii} = k_i$ and
$L_{ij} = -A_{ij}$ for $i \neq j$. Solving Eq.~\ref{eq:lap} in terms
of the normal modes $\phi_i (t)$, we obtain
\begin{equation}
\phi_i (t) = \phi_i (0) ~{\rm
e}^{-\lambda_i t},~~i=1, \ldots, N,
\end{equation}
where $\lambda_i$ are the eigenvalues of {\bf L$^{\prime}$ = D$^{-1}$
L} with {\bf D} being a diagonal matrix having $D_{ii} = k_i$. All the
eigenvalues of {\bf L$^{\prime}$} are real as the matrix is related to
the symmetric normalized Laplacian {\bf D$^{1/2}$ L$^{\prime}$
D$^{-1/2}$} through a similarity transformation. 
Any difference in time-scales of the different modes is manifested as
gaps in the spectrum of this Laplacian, and reveals different levels
of mesoscopic organization in the network. The mode corresponding to
the smallest eigenvalue is associated with global synchronization.
The other modes provide information about collective behavior
within different groups of oscillators. Typically, the size of the
clusters considered decrease with increasing value of the
corresponding eigenvalues.

\begin{figure*}
\begin{center}
\includegraphics[width=0.99\textwidth]{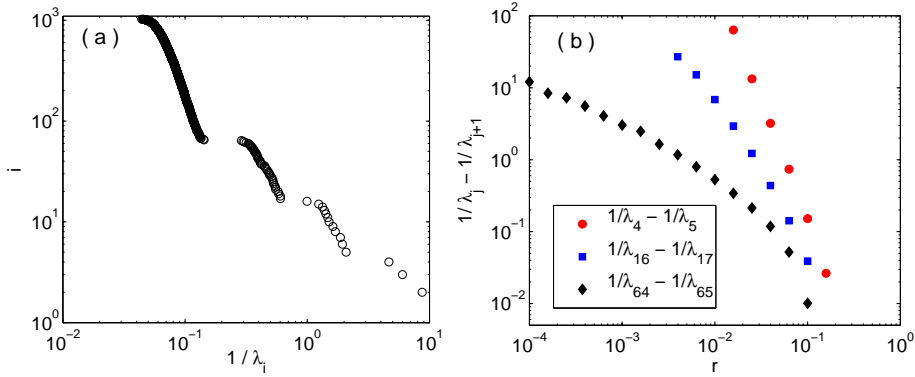}
\caption{(a) The rank-ordered reciprocal eigenvalues for the Laplacian
matrix of a hierarchical modular random network ($h_{lev} = 3$) with 
$N=1024$ oscillators having $\langle k \rangle = 14$ connections (on
average)
for $r =0.05$. The existence of three distinct 
spectral gaps reflects the
three hierarchical levels in which the modular organization is
arranged.
(b) The width of the three Laplacian spectral gaps shown as a function 
of $r$ for a hierarchical modular network with $N=1024$ oscillators 
having $\langle k \rangle = 60$ connections (on average) arranged into
$h_{lev} = 3$ levels. The spectral gap for each level increases with
decreasing $r$, implying that as the hierarchical modular character of
the network becomes more prominent, the synchronization times for the different
levels deviate significantly.
The divergence of the gaps for sufficiently low values of $r$
indicate that the system may not synchronize
globally for very weak coupling among the clusters of modules. 
For all cases, the branching at each level is $q = 4$.}
\label{fig:fig4}
\end{center}
\end{figure*}
In Fig.~\ref{fig:fig4}~(a) we observe multiple gaps in the Laplacian
spectrum for a hierarchical network, the number of gaps corresponding
exactly to the number of hierarchical levels ($h_{lev} = 3$)
indicating that the existence of several distinct time-scales for the
dynamical system has its origin in the hierarchical modular nature. As
seen in Fig.~\ref{fig:fig4}~(b), the width of the gaps increase with
decreasing value of $r$, suggesting that the time-scales become more
differentiated with increasing modularity, i.e., decreasing
connectivity between the modules, between the meta-modules, and so on.

\section{Discussion}
In this paper we have generalized our earlier result for time-scale
separation of dynamics on simple modular networks by showing that for a
system with 
hierarchically nested modules occurring at multiple levels, there
exists a number of distinct time-scales (equal to the number of
hierarchical levels). Any collective dynamical process taking place
on such a system (e.g., coordination of nodal activity) 
will occur at many different temporal rates,
the fastest events taking place at the lowest modular level and the
slowest occurring at the global level. An important point to note is
that the non-uniform synchronization observed in our system is rooted
in its non-trivial
mesoscopic organization. This distinguishes the behavior
from the ``hierarchical" synchronization observed in
homogeneous scale-free networks, where oscillators having more
connections (i.e., higher degree) show stronger phase-locking between
them~\cite{Zhou06}. Thus, the latter phenomenon is related to the dispersion of
microscopic properties (viz., degree) of the nodes, which by design is
identical for all oscillators in our system. An interesting topic for
future study is the effect of different
varieties of degree distribution on synchronization dynamics in
hierarchical modular networks.

Our results also have ramifications for the evolution of such
mesoscopic organization in natural systems such as the brain. While
synchronization of activity among elements belonging to the same
cluster may have functional relevance, global synchronization of the
system is often pathological and detrimental to the
system~\cite{Lumer91,Lumer92}. Thus, the
occurrence of a hierarchically nested modular organization may be
nature's way to allow rapid local synchronization at small scales while
making it difficult for activity in the system as a whole to get
coordinated. 
Note that this possible functional role of hierarchical modularity in
the brain is distinct from the recent suggestion that it is related to
the task of simultaneously segregating sensory signals and
integrating information from multiple channels in the
brain, as the latter requires the existence of hubs (i.e., nodes
having a much higher number of connections than the average, which by
design is absent in our model) in
addition to the modular and
hierarchical nature of cortico-cortical connections~\cite{ZamoraLopez10}.
We have earlier demonstrated that the occurrence of 
modularity in the {\em
C. elegans} nervous system may be directly related to the functional
requirement of 
information processing~\cite{Pan10}. A recent study of
Hopfield-like associative memory models has also shown that modular
network structure alters the attractor landscape of convergence
dynamics in such systems, making the recall of stored patterns more
efficient~\cite{Pradhan11}. Generalizing these results to modules at
more than one level should bring us closer to answering the question
as to
why hierarchical modularity is ubiquitous in the biological world.

\acknowledgments
We would like to thank R. K. Pan for his assistance with the
simulations and R. E. Amritkar for helpful discussions.

\end{document}